\documentclass[onecolumn,floatfix,superscriptaddress,showpacs,showkeys,nofootinbib,preprint]{revtex4-1}

\usepackage{epsfig}
\usepackage{amssymb,latexsym,amsmath}
\usepackage{color}
 
\begin{document}

\title{Some Considerations concerning the Fluctuation \\of the Ratios of two Observables}
\author{M. Hauer}
\affiliation{Frankfurt Institute for Advanced Studies, Frankfurt am Main, Germany}
\affiliation{Institut f\"ur Theoretische Physik, Goethe-Universit\"at, Frankfurt am Main, Germany}
\author{S. Vogel}
\affiliation{SUBATECH,
Laboratoire de Physique Subatomique et des Technologies Associ\'ees \\
University of Nantes - IN2P3/CNRS - Ecole des Mines de Nantes \\
4 rue Alfred Kastler, F-44072 Nantes Cedex 03, France}

\begin{abstract}
We discuss several possible caveats which arise with the interpretation of measurements of fluctuations in heavy ion collisions. 
We especially focus on the ratios of particle yields, which have been advocated as a possible signature of a critical point in the 
QCD phase diagram. We conclude that current experimental observables are not well-defined and are without a proper quantitative meaning.
\end{abstract}



\maketitle

\section{Introduction}
\label{Sec_Intro}

The investigation of fluctuations and correlations, i.e. an event-by-event analysis, of high energy heavy ion collision data has 
been ongoing for several years~\cite{Bleicher:1998wu,Appelshauser:1999ft,Afanasev:2000fu,Aggarwal:2001aa,Adams:2004gp,Chai:2005fj,Mitchell:2005ds,Tarnowsky:2010qp,Aggarwal:2010rf,Tarnowsky:2010zz,Stephanov:1998dy,Stephanov:1999zu,Mishustin:1998eq,Jeon:2003gk}.  
Recently, this discussion was 
refreshed~\cite{Stephanov:2009ra,Koch:2009zz,Konchakovski:2010fh,Kapusta:2010ke,Athanasiou:2010vi,Fu:2010ay,Athanasiou:2010kw} 
in the view that (event-by-event) fluctuations of multiplicity, or of ratios of multiplicities, might serve as an observable to pin down
the critical behavior of the collective, and potentially at some point of its evolution thermalized, system created 
when two heavy ions collide. 
Thus, the measurement of ratio fluctuations might make it possible to fix the values of temperature and baryon chemical potential of a 
critical point (see e.g.~\cite{deForcrand:2003bz,Langelage:2009jb}) in the phase diagram of Quantum Chromo Dynamics (QCD).
In this paper we discuss the problems and possible caveats of such an analysis.

An event-by-event analysis ultimately deals with (joint) distributions of a certain set of observables.
Parameters such as mean value, variance, skewness and kurtosis can be employed to describe the shape
of a distribution obtained. Their values, as the distribution obtained itself, will depend on many experimental factors. 
More complicated measures are often lacking a clear interpretation.
In particular we will investigate the commonly used measure of dynamical ratio fluctuations.

It is not the aim of this paper to suggest an improved observable or method capable of 
locating a critical point of strongly interacting matter, or one that could potentially 
be indicative of new physics.
The goal of this paper is to discuss caveats of the existing methods which are usually not mentioned in the literature. 
The manuscript is organized as follows.
In Section~\ref{Sec_GeneralCons} we formulate some rather general statements about the 
particularities of an event-by-event analysis. 
In Section~\ref{Sec_RatioFluc} we then focus on a particular method and 
discuss some difficulties one might face with its interpretation. 
A Discussion of specific experimental and theoretical studies can be found in Section~\ref{Sec_Discussion}.
Some concluding remarks are summarized in Section~\ref{Sec_Summary}.

\section{General Considerations}
\label{Sec_GeneralCons}

Let us start by making a rather general observation.
The statistical properties of a sample of events depend on the rules chosen to
select this sample from an even larger sample of events, and on the degree of completeness of
the information available about each member of this sample.
In the context of heavy ion collision physics these two aspects translate into
centrality class construction from minimum bias data, and an experiments capabilities of 
particle identification and momentum measurement. We will elaborate on these aspects in the following.

\subsection{Centrality Selection}

We first discuss centrality selection. Two heavy ions collide with relativistic momenta.
Being extended objects, they can do so in a variety of different ways. 
Roughly the following rule should apply (in the average sense): 
The larger the interaction region, the more particles of each species are produced. 
However, the initial state of the collision cannot be observed directly. 
All that can be observed is the final state. 
From this one can then infer the likelihood of a certain initial state. 
Yet, each single possible final state observable will generally suggest a slightly different initial state. 
Hence, the need arises to average over centrality classes. 
The problem is then: Within any such a centrality class will be
events with rather different initial states, altering the true correlation between two observables.
Results will depend on, in particular, which trigger was chosen to construct these centrality 
classes, or sub-samples, of events.

\begin{figure}[ht!]
  \epsfig{file=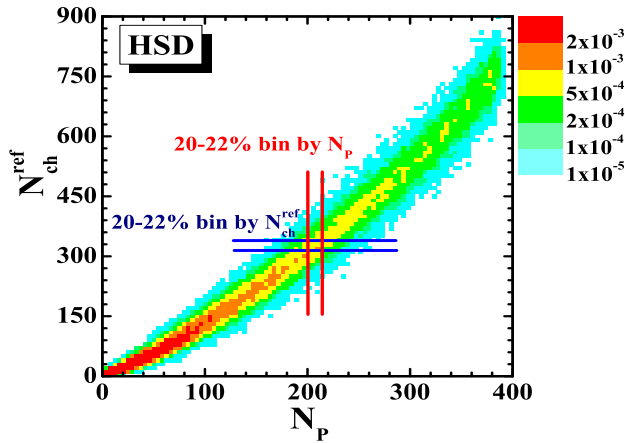,width=8.4cm,height=6.5cm}
  \caption{(Color online)
    Illustration of centrality selection being done in two different ways, once via the number of charged particles in the 
    final states ($N_{ch}^{ref}$) and once via the number of wounded nucleons ($N_P$). 
    This leads to two different sub-samples, despite the fact that the same centrality class ($20-22\%$)
    was constructed. The different colors depict the probability. The figure is taken from \cite{Konchakovski:2008cf}. 
  }
  \label{HSD_centrality}
\end{figure}

To illustrate this point we refer to Fig.~\ref{HSD_centrality}. Shown is a result of a HSD transport simulation, taken 
from~\cite{Konchakovski:2008cf}. 
The two observables shown in Fig.~\ref{HSD_centrality} are the number of participating (or wounded) nucleons~$N_P$,
and the multiplicity of charged final state hadrons~$N_{ch}^{ref}$ in a mid-rapidity acceptance window. 
This analysis is somewhat hypothetical\footnote{As the number of participating nucleons is not accessible to experimental measurement.}, 
but serves to illustrate a crucial point: 
Depending on which one of the two observables is used to define the sample of~$20-22\%$ most central events, 
two rather different sub-samples, indicated by the vertical and horizontal lines, are created. 
These two samples could possibly have rather different statistical properties. 
The two methods are hence not equivalent, and both introduce their very own bias. 
It deserves to be stressed that the smaller the centrality class intervals (here~$2\%$) are chosen the more
distinct are the sub-samples obtained from two different trigger conditions.

\subsection{Acceptance in Momentum Space}

The second aspect we want to discuss is particle acceptance. 
By this we mean particle identification, momentum measurement,
and ultimately the geometric coverage of the interaction region by the detector.
We ignore for now centrality selection and think of a perfectly prepared initial state,  
and explore two limiting cases. The first one being the ideal detector. 
All final state particles are observed, and any correlation can be measured to any degree. 
The opposite limit would be a very bad detector, capable of only detecting a particle every once in a while. 
Such a detector could surely measure the ratios of the occurrence of particles of different species 
(provided sufficient particle identification capabilities) and hence allows for a comparison 
of model results for ratios of averaged particle multiplicities to actual 
data~\cite{BraunMunzinger:1994xr,BraunMunzinger:1995bp,BraunMunzinger:2001ip}.  
However it would be completely unable to inform us how the multiplicities of different particle species 
within one event are correlated or how this correlation would change in different segments of momentum space.
Any realistic detector is in between these limits.

Also here Fig.~\ref{HSD_centrality} can serve as an illustration. 
Changing the acceptance window for charged final state hadrons~$N_{ch}^{ref}$, 
which was for the purpose of this analysis located around mid-rapidity 
(while assuming that~$N_P$ can still be measured with the same accuracy), 
will change the marginal distribution~$P(N_{ch}^{ref})=\sum_{N_P} P(N_P,N_{ch}^{ref})$.
In particular mean value, variance, skewness, and kurtosis of the marginal distribution~$P(N_{ch}^{ref})$
are affected. 
The distribution~$P(N_{ch}^{ref})$ is often used as a reference distribution for construction
of centrality classes by collider experiments. 
That means, depending on the value~$N_{ch}^{ref}$ of a particular event, 
one assigns this event to a certain centrality class. 

\section{Fluctuations of the Ratio of two Observables}
\label{Sec_RatioFluc}

Above considerations served to emphasize the importance of the way by which a certain data set was constructed 
and the degree of completeness of the information available about each element of ones data set. 
These two aspects also determine whether two data sets can really be compared to each other. 
We now focus on one method, the measurement of the fluctuations of the ratio of particle multiplicities, while
leaving it to the reader to decide whether or not some aspects of this discussion could be shared by other methods.

\subsection{Data Distribution}

Firstly a plain observation: 
independent of a particular analysis technique, what is considered here is the ratio of two observables. 
Each of which will have a certain value in each event.
Directly (as done e.g. in the STAR experiment~\cite{Abelev:2009if}) or indirectly (as done e.g. in the NA49 
experiment~\cite{Alt:2008ca}) one can obtain then the ratio of these two observables. 
In the following we will focus on the numbers of particles of a certain species in each event.

We again consider construction of centrality classes and the window of particle acceptance available.
In general one can conclude that the joint (2 dimensional) distribution~$P_{data}(N_1,N_2)$ of the 
multiplicities of the two species '$1$' and '$2$' of particles will be determined by the choices of solutions 
to the two aforementioned aspects, i.e. acceptance and centrality selection.
The distribution of ratios,~$P_{data}(N_1/N_2)$, being constructed from the joint distribution, 
is then certainly no exception. 

A measure for quantifying the width of the distribution of ratios~$P_{data}(N_1/N_2)$, 
similar to the ones used by the STAR \cite{Abelev:2009if} and NA49 collaboration \cite{Alt:2008ca}, could read:
\begin{equation}\label{themeasure}
\sigma^2
~=~ \frac{\langle \left( \Delta N_1 \right)^2 \rangle}{\langle N_1 \rangle^2}
~+~ \frac{\langle \left( \Delta N_2 \right)^2 \rangle}{\langle N_2 \rangle^2}
~-~ 2~\frac{\langle \Delta N_1\Delta N_2 \rangle}{\langle N_1 \rangle \langle N_2 \rangle}~.
\end{equation}

Here~$\langle \left( \Delta N_i \right)^2 \rangle$ denotes the variance of observable~$N_i$,~$\langle N_i \rangle$ denotes 
the mean value of the observable~$N_i$, and~$\langle \Delta N_1\Delta N_2 \rangle$ denotes the co-variance of the observables~$N_1$ 
and~$N_2$.

The above discussion about centrality selection and particle acceptance certainly does not
do justice to all the technical details which will ultimately determine the shape of the
joint distribution~$P_{data}(N_1,N_2)$ of the two observables~$N_1$ and~$N_2$. Yet, 
all five terms, which make up the measure depend strongly on many of these aspects.
Hence, there is no reason to believe that the measure~$\sigma^2$, given in Eq.~\ref{themeasure}, would not depend on them.  

\subsection{Mixed Event Background}

The need arises then to remove the bias introduced. 
The mixed event background subtraction should remove all ``trivial'' correlations due to 
acceptance and centrality selection, while leaving all ``physical'' correlation unchanged. 
Note, that different mixing procedures have been suggested~\cite{Torrieri:2010te}, but since they do not change the 
argumentation of our analysis we do not comment on them separately.  
Before focusing on the mixed event background, let us make another observations about the measure~$\sigma^2$. 
This measure, being so general, could be applied to any sample, as long as the observables~$N_1,N_2$ 
are positive integers.  
Independent of what~$N_1$ and~$N_2$ may represent. 
This implies that also the event mixing procedure, once it is specified, 
could be applied to a more general (or plainly different\footnote{An example, unrelated to physics, could come to mind: 
The number of kindergartens and supermarkets in every city and town could be evaluated. Groups of cities sorted according 
to some criterion, like their estimated number of citizens, would form centrality classes.
Our ability to accurately determine the number of kindergartens and supermarkets would correspond to particle acceptance. 
This might not be a particular thoughtful example, yet it shows that such an analysis is not restricted to 
heavy ion physics. And one still faces the same problem of having to give interpretation to the fluctuation 
measures Eqs.~\ref{themeasure} and~\ref{thedynamicalmeasure}.}) system with observables~$N_1$ and~$N_2$. 

Let us consider a particular event mixing procedure similar to the one used by the NA49 collaboration. 
Firstly a reference distribution~$P_{data}(N)$ of the observable~$N=N_1+N_2$ is constructed: 
\begin{equation}
P_{data}(N)~=~\sum \limits_{N_1=0}^{N} ~P_{data}(N_1,N-N_1)~.
\end{equation}

A mixed distribution~$P_{mix}(N_1,N_2)$ can be obtained by sampling the reference distribution~$P_{data}(N)$
and assigning the values for~$N_1$ and~$N_2$ according to a binomial distribution. 
The mixed event background distribution is then:
\begin{equation}
P_{mix}(N_1,N_2)~=~ q^{N_1} ~(1-q)^{N_2}~\binom{N_1+N_2}{N_2} P_{data}(N_1+N_2)~,
\end{equation}
where~$q = \langle N_1 \rangle / (\langle N_1 \rangle + \langle N_2 \rangle)$ is the probability that a 
randomly drawn particle is of species~'$1$'. 
The distributions~$P_{data}(N)$ and~$P_{mix}(N)$ are then identical.

Stated differently, once the distribution~$P_{data}(N_1,N_2)$ is chosen (or fixed), then also the new mixed background 
$P_{mix}(N_1,N_2)$ is similarly determined. This is to say, for any generic distribution~$P(N_1,N_2)$
a mixed background and the measure of dynamical ratio fluctuations
\begin{equation}\label{thedynamicalmeasure}
\sigma^2_{dyn}~=~ \sigma^2_{data}~-~\sigma^2_{mix}
\end{equation}
are equally determined. 
This is independent of what~$N_1$ and~$N_2$ represent, and independent of how the original data set came about.

\subsection{Illustrations}

We are considering the example of a Bi-variate Normal Distribution (BND),
with five parameters: marginal variances~$\langle \left( \Delta N_1 \right)^2 \rangle_{data}$, and  
$\langle \left( \Delta N_2 \right)^2 \rangle_{data}$, their co-variance~$\langle \Delta N_1\Delta N_2 \rangle_{data}$,
and two mean values,~$\langle N_1 \rangle_{data}$, and~$\langle N_2 \rangle_{data}$. 
Having specified these values, and having decided which event-mixing procedure should be applied, 
the measures,~$\sigma^2_{data}$,~$\sigma^2_{mix}$, and~$\sigma^2_{dyn}$, 
can be straightforwardly obtained.
This implies the existence of a four-dimensional hyper-surface in this parameter space representing 
BNDs with an equal amount~$\sigma^2_{dyn}$ of dynamical ratio fluctuations.

\begin{figure}[ht!]
  \epsfig{file=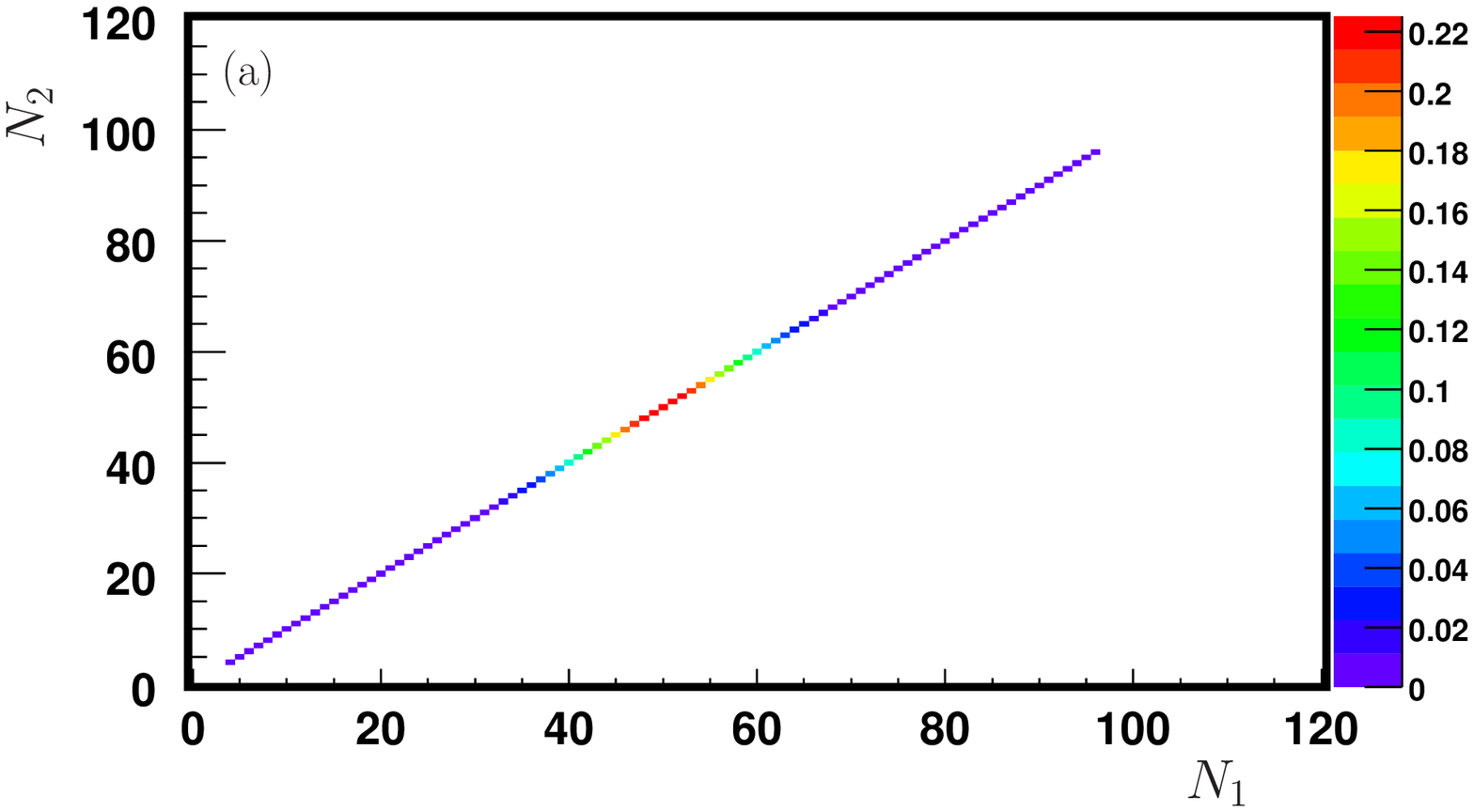,width=8.4cm,height=6.5cm}
  \epsfig{file=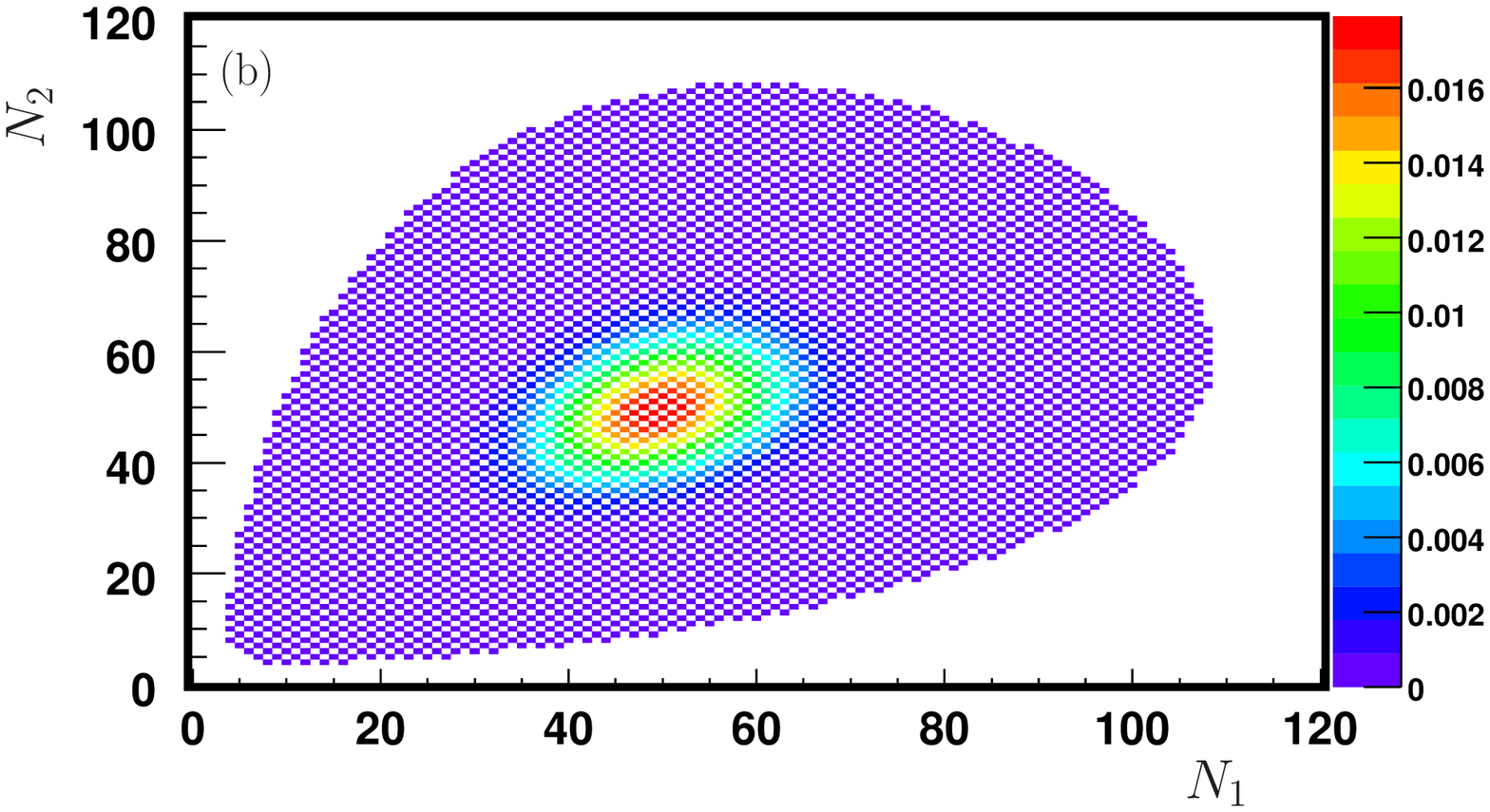,width=8.4cm,height=6.5cm}
  \caption{(Color online)
    Illustration of a perfectly correlated data distribution~({\it left}).
    The mixed event background~({\it right}) shows a wider distribution. The measured
    dynamical ratio fluctuations associated with the original distribution are then negative.
  }
  \label{mixed_event_correlations}
\end{figure}

Let us explore the limit of a BND, Fig.~\ref{mixed_event_correlations}~({\it left}), with completely correlated 
observables~$N_1$ and~$N_2$.
The correlation coefficient 
\begin{equation}
\rho = \frac{\langle \Delta N_1\Delta N_2 \rangle}{\sqrt{\langle \left( \Delta N_1 \right)^2 \rangle \cdot \langle \left( \Delta N_2 \right)^2 \rangle}}
\end{equation}
between them is then~$\rho_{data}(N_1,N_2)=1$. 
The ratio of~$N_1$ and~$N_2$ here in this example is then always equal to unity. 
The same will not be true for the mixed event background, Fig.~\ref{mixed_event_correlations}~({\it right}), 
where the correlation will be weaker,~$\rho_{mix}(N_1,N_2)<1$. 
Following above event mixing procedure, one then quickly finds~$\sigma^2_{data}=0$,~$\sigma^2_{mix}>0$, 
and~$\sigma^2_{dyn}<0$.
In words, despite the fact that the ratio is always one, the measure, discussed in Eqs.~\ref{themeasure} and~\ref{thedynamicalmeasure}, 
suggests a degree of negative dynamical ratio fluctuations for this distribution.

\begin{figure}[ht!]
  \epsfig{file=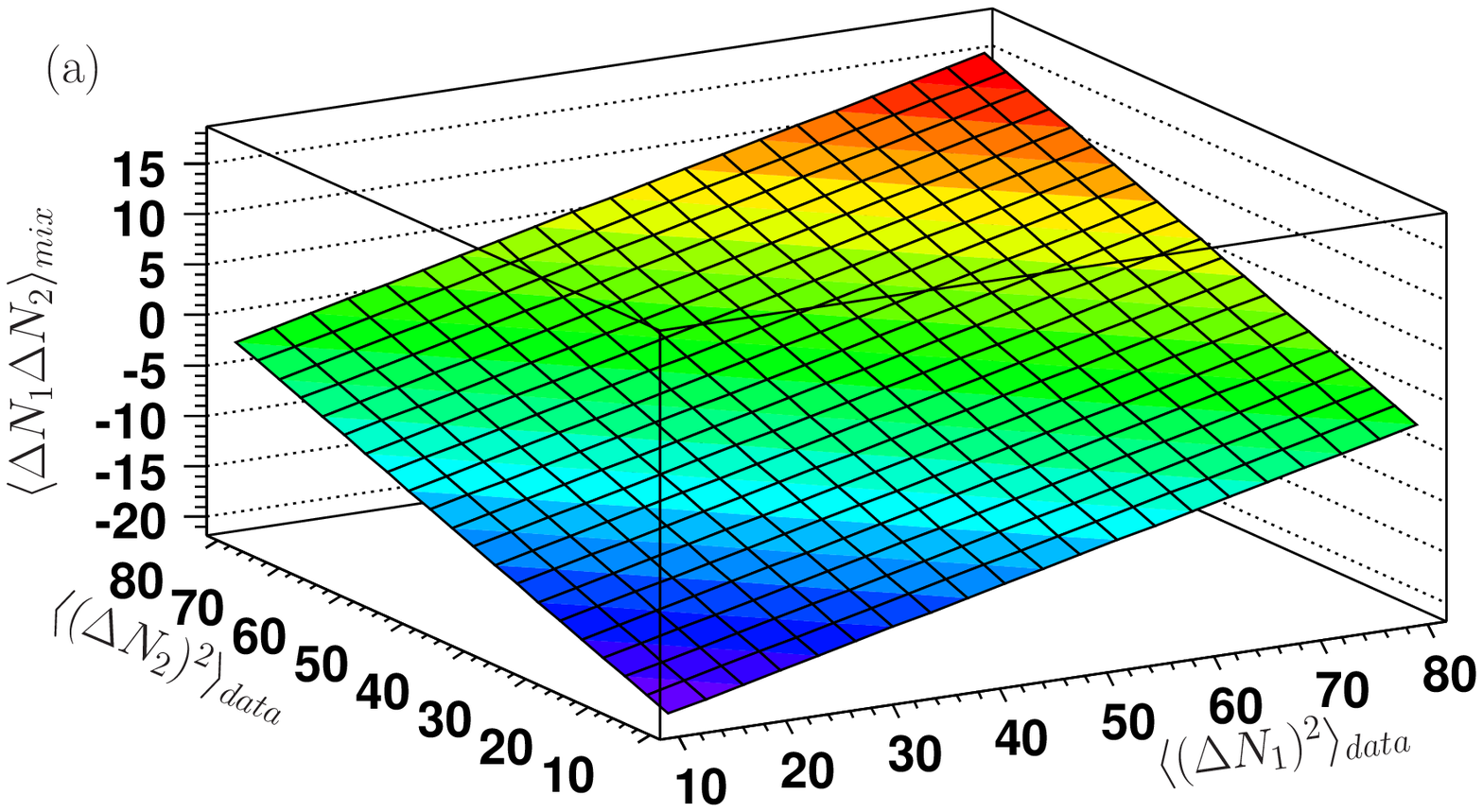,width=8.4cm,height=6.5cm}
  \epsfig{file=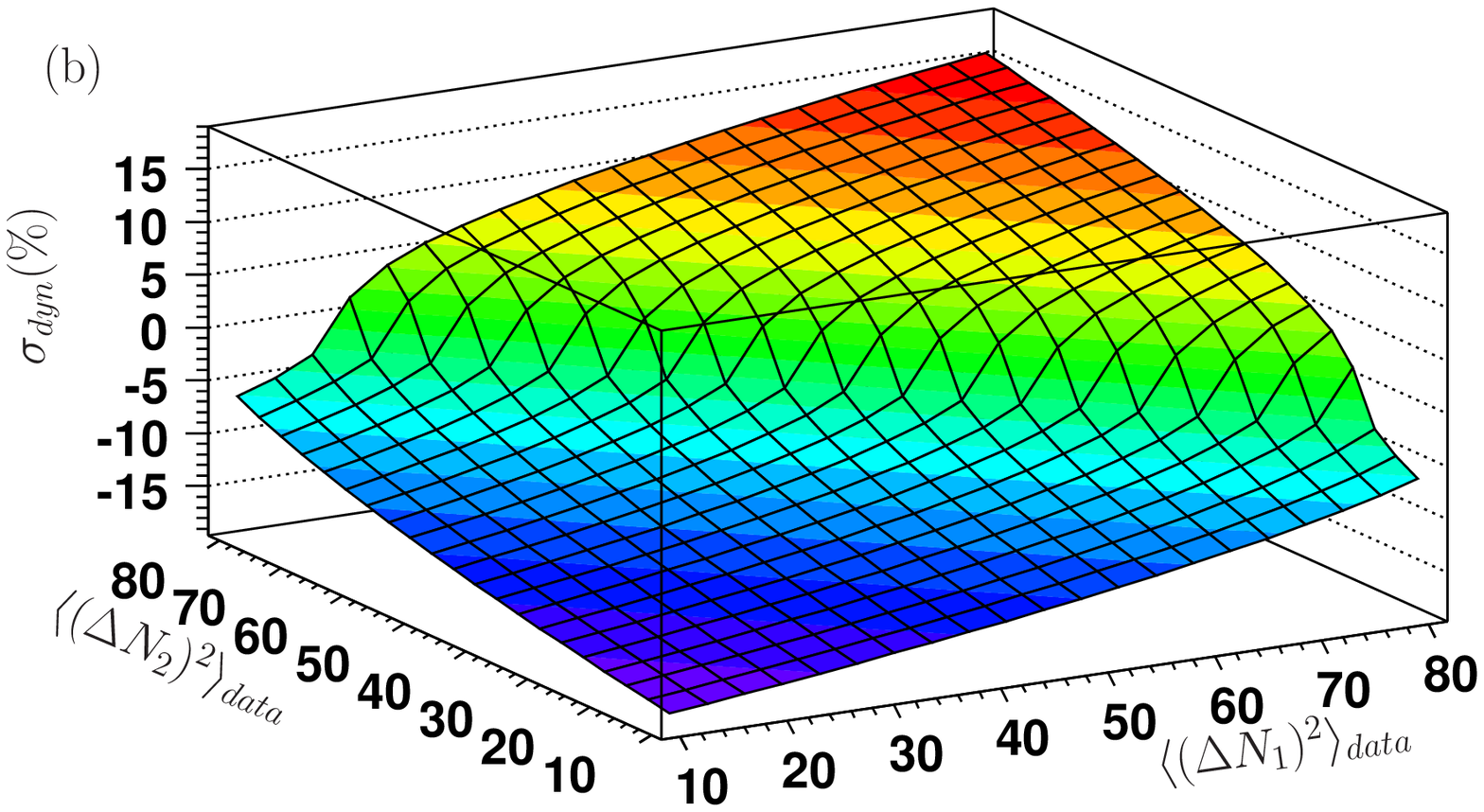,width=8.4cm,height=6.5cm}
  \caption{(Color online)
    Illustration of uncorrelated ($\langle \Delta N_1\Delta N_2 \rangle_{data}=0$) data distribution leading to a 
    correlated ($\langle \Delta N_1\Delta N_2 \rangle_{mix}\not=0$) mixed background~({\it left})
    and resulting dynamic ratio fluctuations~$\sigma_{dyn}$~({\it right}).
  }
  \label{3d_correlations}
\end{figure}

Also other limiting cases could be explored. Here we mention the class
of BNDs with co-variance~$\langle \Delta N_1\Delta N_2 \rangle=0$, as illustrated in Fig.~\ref{3d_correlations}.
The distribution~$P_{data}(N_1,N_2)$ then factorizes into the product 
of two Normal distributions~$P_{data}(N_1,N_2)=P_{data}(N_1)P_{data}(N_2)$.
For the example shown in Fig.~\ref{3d_correlations}, a BND with mean values~$\langle N_1\rangle_{data}=\langle N_2\rangle_{data}=50$, 
and co-variance~$\langle \Delta N_1\Delta N_2 \rangle_{data}=0$ was chosen. 
In Fig.~\ref{3d_correlations} the resulting co-variance of the mixed event 
background~$\langle \Delta N_1\Delta N_2 \rangle_{mix}$~({\it left})
and the measure\footnote{Following the convention we present this number 
as~$\sigma_{dyn} = \textrm{sign}(\sigma_{dyn}^2) ~\sqrt{|\sigma_{dyn}^2|} \cdot 100 ~\%$.} 
of dynamical ratio fluctuations~$\sigma_{dyn}$~({\it right}) are shown in their dependence on the 
variances~$\langle \left( \Delta N_1 \right)^2 \rangle_{data}$ and~$\langle \left( \Delta N_2 \right)^2 \rangle_{data}$.
In general one finds~$\langle \Delta N_1\Delta N_2 \rangle_{mix} \not=0$ after event mixing,
i.e. the mixed event background now contains correlations, which the original sample did not.

\section{Discussion}
\label{Sec_Discussion}

Having formulated rather general statements about the particularities of an event-by-event analysis of
heavy ion collision data, and having discussed a specific fluctuation measure, 
we feel the need to add a few more specific comments.

Measurements of fluctuations and correlations of particle multiplicities are often mentioned with regard to the detection of a 
critical point in the QCD phase diagram or other predictions for the properties of yet to be observed new states of matter. 
Certain aspects of distributions are supposed to shed light on their (temporary) formation. 
Given the caveats discussed throughout this paper the authors are less optimistic. 

The effects discussed above are hard to quantify as this would have to be done specifically and systematically.
The biases introduced by a particular centrality selection criterion and a given particle acceptance 
would have to be studied separately for each observable. 
Focus should be put on the different solutions realized in the two main groups of experimental set-ups, 
fixed target and collider geometries. 
This was not the aim of this paper. Yet some examples can be quoted. 
Here we want to refer to a recent review paper~\cite{Konchakovski:2010fh}, summarizing the PhD thesis of V.~Konchakovski.
In particular we mention Sections 6 and 8.

The first example we want to quote is the correlation coefficient~\cite{Tarnowsky:2007nj,Tarnowsky:2007rr} 
between the multiplicities of charged hadrons in two narrow symmetrically around mid-rapidity arranged acceptance intervals, 
usually termed `forward-backward correlations`.
These correlations have been studied in their dependence on choice of centrality trigger and size of centrality 
classes~\cite{Konchakovski:2008cf,Konchakovski:2010fh,Bzdak:2009xq}. 
A wider definition of centrality classes leads to a stronger correlation between
forward and backward hemispheres. 
A larger system will have lots of particles in both acceptance windows. 
A smaller system will obviously have less particles. 
Combining systems of different sizes into one sample will lead to (strong) correlations in this sample from 
geometric considerations alone. 
A different choice of the centrality trigger will results in a different 
centrality dependence of ones observable.
No unambigous (final state) measure is available to select only events of the `same (initial) system size` into one sample.

For the second example we want to return to the fluctuations of hadron multiplicity ratios.
An HSD simulation of heavy ion collisions has been performed for several center of mass energies. 
At low center of mass energy the data was analyzed twice.
Once for the NA49 acceptance for 3.5\% most central collisions selected via the projectile spectator nucleon signal, 
and once for the STAR acceptance for 5\% most central collisions selected via the reference charged hadron multiplicity distribution.
From above considerations differences between these two set-ups are to be expected.
The situation is summarized as follows~\cite{Konchakovski:2010fh}: 
For some ratios of the multiplicities of selected hadrons stronger differences between the two samples emerge than for others.  
Given our difficulty to interpret the measure of dynamical ratio fluctuations,
we find it hard to make a positive conclusion about the apparent numerical similarity
of the results of the NA49~\cite{:2008ca} and STAR~\cite{Das:2006zg} collaborations.

It is equally hard to disentangle different contributions to ones fluctuation and correlation signals. 
The fact remains that correlation must not be confused with causation. 
We cordially acknowledge the commonly quoted motivation for performing statistical analysis of heavy ion collision data.
However the (conjectured) effects to be investigated, or hoped to be discovered and established, 
are not the only factors determining the statistical properties of ones data sample. 

The view that a (sharp) non-monotonic behavior of a suitably chosen observable evaluated on a suitably constructed 
and modified data set can straightforwardly be taken as evidence for new physics (e.g. the detection of a critical point) is not held by the authors.

\section{Summary}
\label{Sec_Summary}

This discussion largely fell into two parts, each of which tried to raise two points. 
Firstly the results of any event-by-event analysis do depend strongly on how one constructs sub-samples
from minimum bias data. 
Minimum bias data from different experimental set-ups can vary quite strongly, 
and no initial state can be observed.
Secondly, and equally important, is an experiments capability to record accurate and detailed information 
about each event. The patchier the data set the less detailed and conclusive any analysis of it,
or comparison to another data set, can be.  

We then discussed a particular procedure. Also here we tried to raise two points. 
Firstly a lot of information is discarded in favor of a single number,~$\sigma_{dyn}^2$.
This number might not be very indicative of a systems properties. 
Secondly, plainly constructing a reference distribution from a given data distribution according to some set of rules 
does not necessarily help to subtract correlations. 
Unfortunately no reference (data) sample (where a certain effect is simply not contained) is available 
in high energy heavy ion collision physics.

Given above considerations we find it hard to compare and interpret results from different experiments. 
Especially the dependence on the center of mass energy of the colliding ions might then carry little
information.

\begin{acknowledgments}
M.H. acknowledges the financial support received from the Helmholtz International Center 
for FAIR within the framework of the LOEWE program.
We are grateful to Hans Beck, Lorenzo Ferroni, Thorsten Kolleger, Volodymyr Konchakovski, Giorgio Torrieri, Irina Sagert, and Tim Schuster for fruitful discussions.
\end{acknowledgments}



\begin{thebibliography}{100}

\bibitem{Bleicher:1998wu}
  M.~Bleicher {\it et al.},
  Phys.\ Lett.\  B {\bf 435}, 9 (1998).

\bibitem{Appelshauser:1999ft}
  H.~Appelshauser {\it et al.}  [NA49 Collaboration],
  Phys.\ Lett.\  B {\bf 459}, 679 (1999).

\bibitem{Afanasev:2000fu}
  S.~V.~Afanasev {\it et al.}  [NA49 Collaboration],
  Phys.\ Rev.\ Lett.\  {\bf 86}, 1965 (2001).

\bibitem{Aggarwal:2001aa}
  M.~M.~Aggarwal {\it et al.}  [WA98 Collaboration],
  Phys.\ Rev.\  C {\bf 65}, 054912 (2002).

\bibitem{Adams:2004gp}
  J.~Adams {\it et al.} [ STAR Collaboration ],
  J.\ Phys.\ G {\bf G34}, 799-816 (2007).

\bibitem{Chai:2005fj}
  Z.~w.~Chai {\it et al.}  [PHOBOS Collaboration],
  J.\ Phys.\ Conf.\ Ser.\  {\bf 27}, 128 (2005).

\bibitem{Mitchell:2005ds}
  J.~T.~Mitchell,
  J.\ Phys.\ Conf.\ Ser.\  {\bf 27}, 88 (2005).

\bibitem{Tarnowsky:2010qp}
  T.~J.~Tarnowsky,
  J.\ Phys.\ Conf.\ Ser.\  {\bf 230}, 012025 (2010).

\bibitem{Aggarwal:2010rf}
  M.~M.~Aggarwal {\it et al.} [ STAR Collaboration ],
  Phys.\ Rev.\  {\bf C82}, 024912 (2010).

\bibitem{Tarnowsky:2010zz}
  T.~J.~Tarnowsky [ STAR Collaboration ],
  J.\ Phys.\ G {\bf G37}, 094037 (2010).

\bibitem{Stephanov:1998dy}
  M.~A.~Stephanov, K.~Rajagopal and E.~V.~Shuryak,
  Phys.\ Rev.\ Lett.\  {\bf 81}, 4816 (1998).

\bibitem{Stephanov:1999zu}
  M.~A.~Stephanov, K.~Rajagopal and E.~V.~Shuryak,
  Phys.\ Rev.\  D {\bf 60}, 114028 (1999).

\bibitem{Mishustin:1998eq}
  I.~N.~Mishustin,
  Phys.\ Rev.\ Lett.\  {\bf 82}, 4779 (1999).

\bibitem{Jeon:2003gk}
  S.~Jeon and V.~Koch,
  arXiv:hep-ph/0304012.

\bibitem{Stephanov:2009ra}
  M.~A.~Stephanov,
  Phys.\ Rev.\  {\bf D81}, 054012 (2010).

\bibitem{Koch:2009zz}
  V.~Koch,
  PoS {\bf CPOD2009}, 002 (2009).
  
  
\bibitem{Konchakovski:2010fh}
  V.~P.~Konchakovski, M.~I.~Gorenstein, E.~L.~Bratkovskaya and W.~Greiner,
  J.\ Phys.\ G {\bf 37}, 073101 (2010).


\bibitem{Kapusta:2010ke}
  J.~I.~Kapusta,
  Phys.\ Rev.\  {\bf C81}, 055201 (2010).

\bibitem{Athanasiou:2010vi}
  C.~Athanasiou, K.~Rajagopal, M.~Stephanov,
  [arXiv:1008.3385 [hep-ph]].

\bibitem{Fu:2010ay}
  W.~-j.~Fu, Y.~-l.~Wu,
  Phys.\ Rev.\  {\bf D82}, 074013 (2010).

\bibitem{Athanasiou:2010kw}
  C.~Athanasiou, K.~Rajagopal, M.~Stephanov,
  Phys.\ Rev.\  {\bf D82}, 074008 (2010).

\bibitem{deForcrand:2003bz}
  P.~de Forcrand, O.~Philipsen,
  [hep-ph/0301209].

\bibitem{Langelage:2009jb}
  J.~Langelage, O.~Philipsen,
  JHEP {\bf 1001}, 089 (2010).

\bibitem{Konchakovski:2008cf}
  V.~P.~Konchakovski, M.~Hauer, G.~Torrieri {\it et al.},
  Phys.\ Rev.\  {\bf C79}, 034910 (2009).

\bibitem{BraunMunzinger:1994xr}
  P.~Braun-Munzinger, J.~Stachel, J.~P.~Wessels {\it et al.},
  Phys.\ Lett.\  {\bf B344}, 43-48 (1995).

\bibitem{BraunMunzinger:1995bp}
  P.~Braun-Munzinger, J.~Stachel, J.~P.~Wessels {\it et al.},
  Phys.\ Lett.\  {\bf B365}, 1-6 (1996).

\bibitem{BraunMunzinger:2001ip}
  P.~Braun-Munzinger, D.~Magestro, K.~Redlich {\it et al.},
  Phys.\ Lett.\  {\bf B518}, 41-46 (2001).

\bibitem{Abelev:2009if}
  B.~I.~Abelev {\it et al.} [ STAR Collaboration ],
  Phys.\ Rev.\ Lett.\  {\bf 103}, 092301 (2009).

\bibitem{Alt:2008ca}
  C.~Alt {\it et al.} [ NA49 Collaboration ],
  Phys.\ Rev.\  {\bf C79}, 044910 (2009).

\bibitem{Torrieri:2010te}
  G.~Torrieri, R.~Bellwied, C.~Markert {\it et al.},
  J.\ Phys.\ G {\bf G37}, 094016 (2010).
  

\bibitem{Tarnowsky:2007nj}
  T.~Tarnowsky, R.~Scharenberg and B.~Srivastava,
  Int.\ J.\ Mod.\ Phys.\  E {\bf 16}, 1859 (2007).

\bibitem{Tarnowsky:2007rr}
  T.~J.~Tarnowsky  [STAR Collaboration],
  PoS C {\bf POD07}, 019 (2007).
 
\bibitem{Bzdak:2009xq}
  A.~Bzdak,
  Phys.\ Rev.\  C {\bf 80}, 024906 (2009).

\bibitem{:2008ca}
  C.~Alt {\it et al.}  [NA49 Collaboration],
  Phys.\ Rev.\  C {\bf 79}, 044910 (2009).

\bibitem{Das:2006zg}
  S.~Das  [STAR Collaboration],
  J.\ Phys.\ G {\bf 32}, S541 (2006).


\end{thebibliography}
\end{document}